\documentclass[sigconf]{acmart}
\AtBeginDocument{%
  \providecommand\BibTeX{{%
    \normalfont B\kern-0.5em{\scshape i\kern-0.25em b}\kern-0.8em\TeX}}}

\setcopyright{acmlicensed}
\copyrightyear{2024}
\acmYear{2024}
\acmDOI{XXXXXXX.XXXXXXX}

\usepackage{tikz}

\usepackage{algorithm}
\usepackage{algpseudocode}
\usepackage{makecell}

\graphicspath{{./Figures/}}

\begin{document}

\title{Optimal Control-Based Falsification of Learnt Dynamics via Neural ODEs and Symbolic Regression}


\author{Lasse Kötz}
\email{kotz@chalmers.se}
\affiliation{
\institution{Chalmers University of Technology}
\city{Gothenburg}
\country{Sweden}
}

\author{Jonas Sjöberg}
\affiliation{
\institution{Chalmers University of Technology}
\city{Gothenburg}
\country{Sweden}
}

\author{Knut Åkesson}
\email{knut@chalmers.se}
\affiliation{
\institution{Chalmers University of Technology}
\city{Gothenburg}
\country{Sweden}
}

\keywords{Falsification, Optimal control, Surrogate modeling, Neural ODEs, Symbolic Regression}

\begin{abstract}
We present a falsification framework that integrates learned surrogate dynamics with optimal control to efficiently generate counterexamples for cyber-physical systems specified in signal temporal logic (STL). The unknown system dynamics are identified using neural ODEs, while known a-priori structure is embedded directly into the model, reducing data requirements. The learned neural ODE is converted into an analytical form via symbolic regression, enabling fast and interpretable trajectory optimization. Falsification is cast as minimizing STL robustness over input trajectories; negative robustness yields candidate counterexamples, which are validated on the original system. Spurious traces are iteratively used to refine the surrogate, while true counterexamples are returned as final results. Experiments on ARCH-COMP 2024 benchmarks show that this method requires orders of magnitude fewer experiments of the system under test than optimization-based approaches that do not model system dynamics.

\end{abstract}
\maketitle

\section{INTRODUCTION}
\begin{figure}
    {

\definecolor{green}{RGB}{40,188,0}
\definecolor{red}{RGB}{155,0,0}
\definecolor{darkgreen}{RGB}{8,137,0}
\definecolor{darkblue}{RGB}{9,52,226}
\definecolor{state3background}{RGB}{180, 185, 193}

\usetikzlibrary{arrows,automata,calc,positioning, shapes.geometric}

\tikzset{
    state2/.style={
           rectangle,
           draw=black, very thick,
           minimum height=2em,
           inner sep=2pt,
           minimum size=3em,
           font=\huge
           },
    state4/.style={
           diamond,
           draw=black, very thick,
           minimum height=1em,
           inner sep=1pt,
           minimum size=1em,
           font=\huge
           }
}

\begin{tikzpicture}[->,>=stealth',shorten >=1pt,auto,semithick, transform shape, scale=0.6]

\node[state2] (1) [scale=1.0]{
    \begin{tabular}{c}
        Initializer
    \end{tabular}};

\node[state2] (2) [right=2.0cm of 1, scale=1.0] {
    \begin{tabular}{c}
        SUT \\
        experiment
    \end{tabular}
    };

\node[state4] (3) [right=2cm of 2, scale=1.0, aspect=2] {
    \begin{tabular}{c}
        Specification \\
        falsified?
    \end{tabular}
    };

\node[state2] (4) [below=1.5cm of 3, scale=1.0] {
    \begin{tabular}{c}
         Neural ODE  \\
         approximator
    \end{tabular}
    };

\node[state2] (5) [left=2cm of 4, scale=1.0] { 
    \begin{tabular}{c}
         Symbolic \\ distillation 
    \end{tabular}
};

\node[state2] (6) [left=2cm of 5, scale=1.0] {
    \begin{tabular}{c}
         Optimal \\ control
    \end{tabular}};

\node[state2] (7) [below=1.5cm of 6, scale=1.0] {
    \begin{tabular}{c}
        Robustness \\
        approximator
    \end{tabular}
};

 





\coordinate[left=1.5cm of 1] (d1);
\coordinate[above=1.5cm of 3] (d2);
\coordinate[below=1.5cm of 7] (d3);
\coordinate[above=1cm of 3] (d4);
 
\path
    (1) edge [above] node[scale=1.0, font=\LARGE] {
        \begin{tabular}{c}
        Inputs \\
        $u(t)$
        \end{tabular}
        } (2)
    (2) edge [above] node[scale=1.0, font=\LARGE] {
        \begin{tabular}{c}
            Outputs \\
            $y(t)$
        \end{tabular}
        } (3)
    (3) edge [right] node[scale=1.0, font=\LARGE] {
        No
    } 
    (4)
    (4) edge [above] node[scale=1.0, font=\LARGE] {
        \begin{tabular}{c}
             Neural \\ 
             surrogate
        \end{tabular}
    } (5)
    (5) edge [above] node[scale=1.0, font=\LARGE] {
         \begin{tabular}{c}
             Symbolic \\
             surrogate
         \end{tabular}
    } (6)
    (d3) edge [left] node[scale=1.0, font=\LARGE] {
        Specification
    } (7)
    (3) edge [right] node[scale=1.0, font=\LARGE] {
        Yes
    } (d4)
    (7) edge [left]
    node[scale=1.0, font=\LARGE]
    {
        \begin{tabular}{c}
            Approximated  \\ specification 
        \end{tabular}
    } (6)
    ;
    
    \draw[->] (6.north) |- (2.west) node[scale=1.0, font=\LARGE, left, pos=0.25] {\begin{tabular}{c}
         Least robust \\
         inputs
    \end{tabular}};

\end{tikzpicture}}
    \caption{The SUT is approximated using neural surrogates with every new trajectory available, before being distilled into a symbolic representation. Counterexamples are computed using optimal control of the symbolic surrogate but must always be verified on the SUT to avoid spurious counterexamples.}
    \label{fig:flow} 
\end{figure}
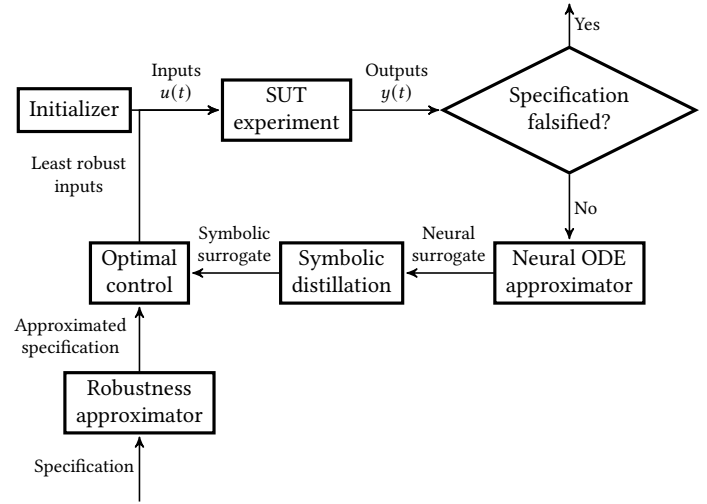
Rapid technological advancements have made systems increasingly complex. Cyber-physical systems (CPS) are now widely deployed in safety-critical products such as autonomous vehicles and healthcare devices, which puts even greater emphasis on ensuring adherence to desired behaviours. Formal verification techniques—including model checking \cite{baier2008principles} and deductive verification \cite{filliatre2011deductive}—have proven effective for verifying the correctness of smaller systems under test (SUT) or sub-systems. However, for sufficiently complex SUTs, formal verification techniques are generally not applicable or feasible. 


Falsification has emerged as an alternative method that aims to find a \textit{counterexample}, violating given safety specifications. A specification is typically expressed in signal temporal logic (STL), an expressive formalism capable of capturing a wide range of signal behaviors. A counterexample corresponds to a trace of input and output signals for the SUT that violates the specification. In classical falsification, this is done by performing experiments on the SUT, treating it as a black-box model and observing its input–output traces. Depending on the specific SUT, experiments can be carried out either through simulation or by conducting physical experimentation. In practice, model-in-the-loop simulations can be performed in early design phases and at a much larger scale than physical experiments, which are typically far more costly and time-consuming to conduct.

When a counterexample is found, it can be used to correct the identified safety violation in the SUT. A counterexample is an input–output trajectory that violates the specified safety property. Falsification avoids many of the challenges associated with formal verification, as it only requires that experiments can be conducted on the SUT—either in simulation or in by physical experiments. However, existing falsification methods typically fail to exploit all information gained from experiments, instead relying on repeatedly running new experiments. 




In this work, we propose an optimal control-based falsification method built on neural ODEs \cite{chen2018neural} and symbolic regression (SR), as shown in Figure \ref{fig:flow}. In short, a neural ODE models the system dynamics by representing the time derivative of the state with a neural network, yielding a nonlinear differential equation whose solution approximates the state evolution. The neural surrogate is then used to distill a symbolic representation through symbolic regression techniques. With this symbolic model available, an optimal control problem is formulated to steer the surrogate toward a trajectory that violates the STL specification, if possible. 



As a final step, the counterexample is validated by running an experiment on the SUT using the optimal control input and evaluating the true robustness value. Note that the robustness value is defined as positive if the specification is satisfied and negative if it is violated. If the SUT experiment also violates the specification, the algorithm terminates and returns the counterexample as the solution to the falsification problem. If not, the trajectory produced from the SUT experiment is added to the dataset used to learn the surrogate dynamics, and the loop is repeated with the expanded dataset. Using neural ODEs as a surrogate framework naturally supports nonlinear dynamics and full differentiability of the state evolution with respect to the input signal.


Below, we outline the following main contributions of this paper: 
\begin{itemize}
    \item A falsification method is proposed that learns system dynamics from experiments using neural and symbolic surrogate models. The falsification step is performed via an optimal control formulation that relies on the symbolic representation of the learned dynamics.

  An approximation of the specification is incorporated into the optimal control problem to guide the search toward violating trajectories. To the authors’ knowledge, this is the first method that combines learning-based nonlinear surrogate modeling with an optimal control approach for falsification.
    \item The performance is evaluated and compared for several benchmark specifications in the ARCH-COMP competition \cite{khandait2024arch}, including performance comparison to the participating tools in the competition.
\end{itemize}

\section{RELATED WORK}
The falsification problem has been approached in many different ways. In black-box falsification, specifications are often violated using heuristic or gradient-free optimization methods, where input generators are parametrized and the optimization procedure attempts to find parameter values that falsify the specification. Successful approaches include, for example, hybrid-corners random, Nelder–Mead, and line-search falsification.

More recent methods aim to learn a surrogate of the objective function—rather than the system dynamics—and balance exploration and exploitation, typically using Bayesian optimization (BO) \cite{shahriari2015taking}. These strategies \cite{deshmukh2017testing,ramezani2025falsification} guide sampling toward regions of the input-parameter space where low robustness is expected, thereby reducing the number of SUT evaluations required for falsification.

However, these methods lack the ability to model the dynamics explicitly. Instead, they rely on representing the surrogate as a simplified probabilistic mapping between input signal parameters and the expected objective value. Furthermore, they scale poorly with the dimensionality of the input parameter space, which further constrains parameterization.

A further step in the development has been to approximate a surrogate model for the underlying dynamical system iteratively, and subject it to optimization-based falsification methods instead of attempting to falsify the relatively expensive SUT directly. This approach offers flexibility but is only effective when the discrepancy between the SUT and the surrogate remains relatively small. Prior tools include ARIsTEO \cite{menghi2020approximation}, which leverages system identification to identify a range of models and subjects the cheaper surrogate to black-box falsification.

A related approach \cite{bak2024falsification} is to use Koopman operator theory to model the time evolution of a set of observables. This provides linear dynamics of the observables via dynamic mode decomposition (DMD), which enables the use of MILP-solvers to find the least robust trace and verifying them on the SUT. For nonlinear dynamics in general, however, there need not be a finite subspace of observables that is invariant under the Koopman operator, making them inherently inexact. Furthermore, DMD struggles with linear approximations for systems with multiple fixed points or periodic orbits, which are common properties of nonlinear dynamical systems.

Other notable surrogate model-based frameworks include NNFal \cite{kundu2024data} which trains a static depth deep neural network as a surrogate, mapping input vectors to output vectors. The falsification/verification step is carried out using adversarial attack algorithms and deep neural network verifiers. It is however, limited in what kind of temporal operators that it supports during falsification and does not learn any continuous-time nonlinear dynamics in which prior knowledge can be embedded. $\Psi -$TaLiRo \cite{thibeault2021psy} is another framework with industrial adaptation that searches for minimal robustness trajectories from arbitrary Simulink and Stateflow diagrams based on stochastic optimization techniques. 

Online diffusion (OD) \cite{peltomaki2022wasserstein} is an online test generation algorithm, based on Wasserstein generative adversarial networks. The algorithm works by denoising test vectors corresponding to parameterizations of piecewise constant input signals. Moonlight \cite{nenzi2023moonlight} is a monitoring tool and falsification framework which in the ARCH-COMP 2024 competition leverages BO with trust regions. ForeSee \cite{zhang2021effective} is a framework which manipulates the specification's syntax tree to avoid the so-called scale problem that arises when falsifying signals of widely different magnitudes \cite{zhang2021effective}. The falsification strategy uses Monte Carlo tree search to traverse the specification syntax tree. In FalCAuN \cite{waga2020falsification} the SUT is treated as a black-box transition system with arbitrary many states. It uses model checking with active automata learning to represent the system as a Mealy machine. EXAM-Net uses an adversarial approach to have the discriminator learn a mapping between inputs and robustness value. ATheNA \cite{formica2023search} is a framework using a mix of manual and automatic fitness functions. The toolbox allows the engineer to target certain areas of the input space which may be especially critical for falsification. 

To summarize, existing frameworks either do not model the full nonlinear dynamics of the system explicitly or do not provide usable gradients for gradient-based falsification of the surrogate, motivating our proposed framework. 
\section{BACKGROUND}
\subsection{Preliminaries}
\begin{definition}[trace]
A trace is defined as a tuple
\begin{equation}
    \tau = (x_0, u(t), y(t))
\end{equation}
consisting of the initial state $x_0$, output signal $y$, and input signal $u$ sampled at discrete time steps $\mathbf{t}$ when obtained from experiments; continuous-time traces are discretized for robustness evaluation. In this paper, a trace can arise from:
\begin{itemize}
    \item solving an optimal control synthesis problem,
    \item solving an initial-value problem (IVP) of the surrogate, or
    \item performing experiments of the SUT to obtain an input-output trajectory.
\end{itemize}
\end{definition}

\begin{definition}[Signal Temporal Logic \cite{maler2004}              ]
    From a set of atomic predicates $\mu$, the syntax of STL is defined as
    \begin{equation}
        \varphi ::=\mu \: | \:  \neg\mu \: | \: \varphi \land\psi \: | \: \varphi \lor \psi \: | \: \square_{[a, b]}\psi \: |  \: \lozenge_{[a, b]}\psi \: | \: \varphi \mathcal{U}_{[a, b]} \psi,
    \end{equation}
    where $a \leq b$ are time-bounds for non-negative scalar values. The $\square_{[a,b]}$ and $\lozenge_{[a,b]}$, denote the \textit{globally} and \textit{finally} operators in the closed interval $[a, b]$, respectively. Lastly, $\mathcal{U}_{[a, b]}$ denotes the \textit{until} operator.
\end{definition}
\begin{definition}[STL Semantics]
    The semantics of STL are recursively defined as 
    \begin{equation*}
        \begin{aligned}
            (y, t) & \vDash \mu \quad  & \iff &\mu(y(t)) > 0 \\
            (y, t) & \vDash \neg \mu \quad  &\iff &\neg((y, t) \vDash \mu) \\
            (y, t) & \vDash \varphi\land\psi & \iff & (y, t) \vDash \varphi \land (y, t) \vDash \psi \\
            (y, t) & \vDash \varphi \lor \psi & \iff & (y,t) \vDash \varphi \lor (y,t) \vDash \psi \\
            (y, t) & \vDash \square_{[a, b]} \varphi  & \iff & \forall t' \in [t + a, t+b], (y, t') \vDash \varphi \\
            (y, t) & \vDash \lozenge_{[a, b]} \varphi & \iff & \exists t' \in [t + a, t+b], (y, t') \vDash \varphi \\
            (y, t) & \vDash \varphi \mathcal{U}_{[a, b]} \psi & \iff & \exists t' \in [t+a, t+b], (y,t') \vDash \psi \\ & & & \land \forall t'' \in [t, t'], (y, t'') \vDash \varphi.
        \end{aligned}
    \end{equation*}
\end{definition}
\begin{definition}[Robustness Semantics]
Finding a counterexample as a solution to an optimization problem requires quantifying how strongly the specification is violated or not violated. This \textit{robustness value} is quantified with the following quantitative semantics \cite{leung2023backpropagation}:
\begin{equation*}
    \begin{aligned}
        (y, t) \vDash \varphi & \iff \rho^\varphi(y, t) > 0 \\
        \rho^{\neg \varphi}(y, t) &= -\rho^\varphi(y, t) \\
        \rho^{\varphi_1 \land \varphi_2}(y, t) & = \text{min}(\rho^{\varphi_1}(y, t), \rho^{\varphi_2}(y, t)) \\
        \rho^{\varphi_1 \lor \varphi_2}(y, t)& = \max(\rho^{\varphi_1}(y, t), \rho^{\varphi_2}(y, t)) \\
        \rho^{\lozenge_{[a, b]}\varphi}(y, t) &= \text{max}_{t' \in [t+a, t+b]}(\rho^\varphi(y, t')) \\
        \rho^{\square_{[a, b]}\varphi}(y, t) &= \text{min}_{t' \in [t+a, t+b]}(\rho^\varphi(y, t'))\\
        \rho^{\varphi_1\mathcal{U}_{[a, b]}\varphi_2}(y, t) &= \text{max}_{t' \in [t+a, t+b]}(\text{min}([\rho^{\varphi_2}(y, t'), \\ &\text{min}_{t'' \in [t+a, t']}(\rho^{\varphi_1}(y, t''))]))
    \end{aligned}
\end{equation*}
for some real-valued function $\rho^\varphi: \mathbb{R}^{n_{signals}} \mapsto \mathbb{R}$. 
\end{definition}

\subsection{Problem statement}
Given a SUT $\bar{\mathcal{S}}$ and a specification of desired behavior $\varphi$, the falsification problem is to find a \textit{counterexample} such that
\begin{equation}
    \bar{\mathcal{S}}(x_0, u(t)) \nvDash \varphi,
\end{equation}
where $x_0$ is the initial state, $\varphi$ is the specification, commonly expressed in signal temporal logic and $u(t): \mathbb{R}_{\geq 0} \mapsto \mathbb{R}^m$ are the inputs to $\bar{\mathcal{S}}$. 
\section{LEARNING-BASED FALSIFICATION}
Our proposed framework, Falsification using optimal Control and Neural surrogates (FalConN) solves the falsification problem by combining neural surrogate modeling techniques, symbolic distillation, and optimal control synthesis. With no prior knowledge of the dynamics, the process begins by running an experiment using a randomized input signal, constrained only by specified upper and lower bounds, to initialize the estimation dataset. For simpler specifications, the initial experiment can be selected using alternative heuristic strategies instead. In general, however, the initial experiments should be designed to persistently excite the system, ensuring that the collected traces are informative for learning the dynamics.

The proposed method's main loop is the iterative training-distillation-optimal control loop (steps 4 through 12 in Algorithm \ref{alg:learning-based falsification}), which terminates when a true falsifying trace is found or the number of iterations has reached a preset experiment budget. 

\begin{algorithm}[t]
\caption{Neural-symbolic falsification with optimal control}
\label{alg:learning-based falsification}
\begin{algorithmic}[1]
    \Statex \textbf{Input:} experiment budget, $\bar{\mathcal{S}}$, $\varphi$ 
    \Statex \textbf{Output:} $u_{OC}$
    \State $k \gets 0$ \Comment{Number of experiments on $\bar{\mathcal{S}}$}
    \State $\mathcal{D}_0 \gets \text{InitializeData}
    (\bar{\mathcal{S}})$
    \State $\rho_{inc} \gets \text{ComputeRobustness}(\bar{\mathcal{S}}, \varphi, \mathcal{D}_0)$
    \While{$k \leq \text{experiment budget}$ \textbf{and} $\rho_{inc} > 0$}
        \State $\hat{\mathcal{S}} \gets \text{EstimateNeuralSurrogate}(\mathcal{D}_k)$
        \State $\hat{\mathcal{S}}_{symb} \gets \text{DistillSymbolic}(\hat{\mathcal{S}}, \mathcal{D}_k)$
            \State $u_{OC} \gets \text{SynthesizeControl}(\hat{\mathcal{S}}_{symb}, \varphi)$
            \State $\tau_{OC} \gets \text{Experiment}(\bar{\mathcal{S}}, u_{OC})$
            \State $\rho_{inc} \gets \text{ComputeRobustness}(\bar{\mathcal{S}}, \varphi, \tau_{OC})$
        \State $\mathcal{D}_{k+1} \gets \mathcal{D}_k \cup \{ \tau_{OC} \} $
        \State $k \gets k+1$
    \EndWhile
    \State \textbf{Return:} $u_{OC}$ 
\end{algorithmic}
\end{algorithm}

For each new trajectory added to the estimation dataset $\mathcal{D}_k$, our method approximates a neural surrogate model as shown in Alg \ref{alg:EstimateUDE}. Specifically, we use neural ODEs for estimation, parameterizing the unknown dynamics with a neural network. The neural ODE formulation allows prior knowledge about the system to be embedded directly into the model, so that only the unknown components of the dynamics are learned. This significantly improves data efficiency when estimating the full dynamics. The surrogate is trained by minimizing the discrepancy between the observed and simulated trajectories. Incorporating prior structural information also improves generalization beyond the estimation data and leads to faster training convergence.

Our algorithm distills a symbolic form from the estimated neural surrogate, according to Algorithm \ref{alg:EstimateSymb}, by employing symbolic regression methods (SR). The resulting symbolic surrogate serves as a basis for the optimal control synthesis problem by enabling the use of sparse nonlinear optimizer (SNOPT \cite{gill2002}).

In the last step, the symbolic surrogate representation and STL specification are jointly encoded as an optimal control problem. This optimal control problem can be solved for a least robust input signal $u_{OC}$ using sequential quadratic programming (SQP) solvers and specifically SNOPT by minimizing the robustness value. The resulting solution $u_{OC}$ is verified by executing the SUT for $u_{OC}$ to avoid returning spurious counterexamples due to surrogate incorrectness. If $u_{OC}$ falsifies the SUT, it is returned as a solution to the falsification problem. If it does not falsify the SUT, however, the experiment is added to the estimation dataset $\mathcal{D}$ used to estimate the surrogate in subsequent iterations. 

Alternatively, one could formulate the optimal control problem directly on the learned neural surrogate using a discretized neural ODE. This would avoid the loss of accuracy introduced by symbolic distillation, but it would also result in a substantially more difficult optimization problem, making it challenging to solve efficiently. A detailed exploration of this alternative is beyond the scope of this paper.


\subsection{Neural ODEs}
\begin{figure}
    \centering
    \includegraphics[width=1.0\linewidth]{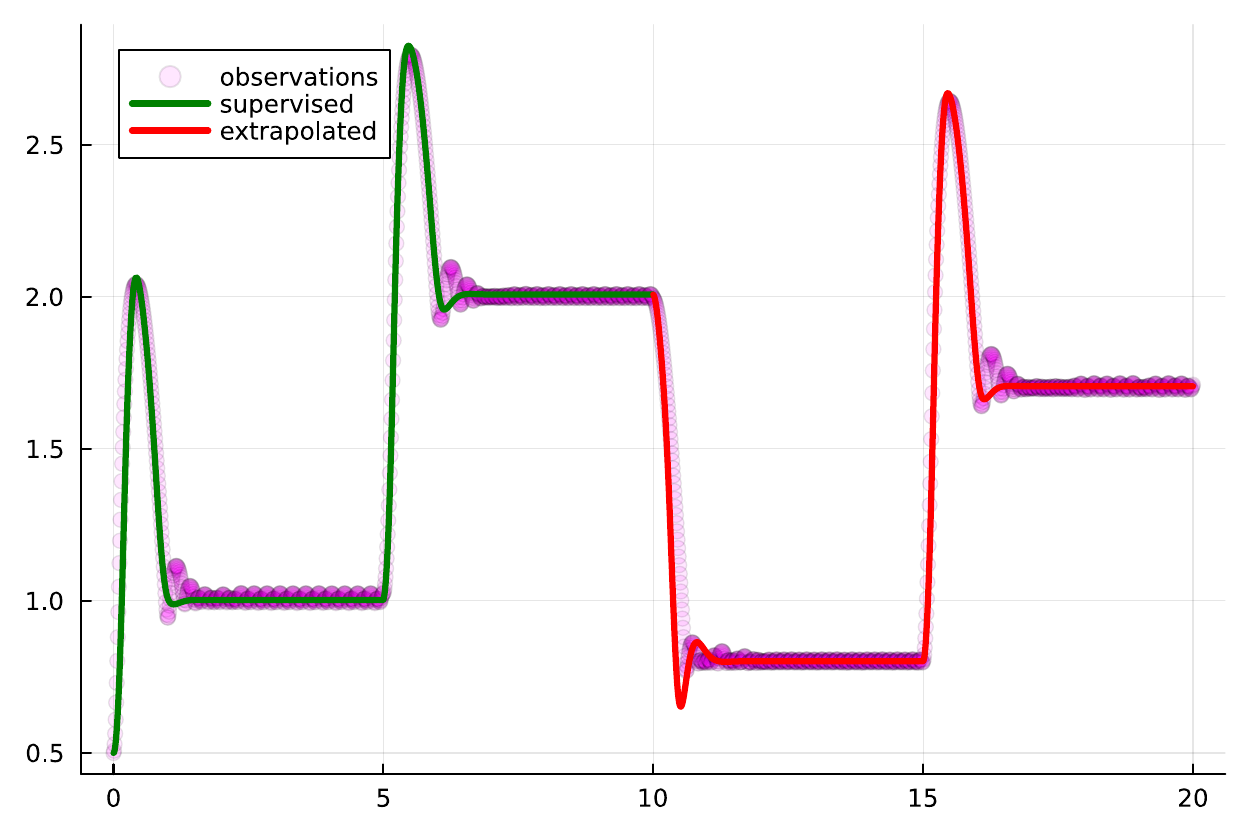}
    \caption{Simulation of neural ODE surrogate trained on estimation data in the interval $t\in [0, 10] $ and extrapolated simulation on unseen data $t \in [10, 20]$. The trajectory is simulated from the NN-controller benchmark example \cite{khandait2024arch}.}
    \label{fig:learned_neuralode}
\end{figure}
We assume that a trajectory can be obtained as the solution to the initial value problem of a 
neural ODE \cite{chen2018neural}, potentially incorporating a priori assumptions of the dynamics
\begin{equation}\label{eq:IVP}
    \begin{aligned}
        \frac{dx}{dt} &= f_{k}(x, u, t) + f_\theta(x, u, t) \\
        x(0) &= x_0,
    \end{aligned}
\end{equation}
where $f_k$ and $f_\theta$ are the known and unknown dynamics, respectively. This is particularly useful when a priori dynamics $f_k$ are available, as it improves generalizability beyond estimation data and reduces estimation time. In the specific case that $f_k = 0$, we are left with a vanilla neural ODE 
\begin{equation}
    x(T) = x_0 + \int_{0}^T  f_\theta(x, u, t) \: dt .
\end{equation}
Neural ODEs can be integrated numerically using any standard ODE-solver, including adaptive step-size solvers, making them analogous to continuous-depth generalizations of residual networks \cite{he2016deep} without fixed time steps. They locally extrapolate well beyond supervised training, as seen in Figure \ref{fig:learned_neuralode}, where trajectories are accurate despite small perturbations in training data.

The unknown dynamics are commonly parameterized as a multi-layer perceptron mapping the vector field $f_\theta: \mathbb{R}^{n \times m} \mapsto \mathbb{R}^n$ for an $n$-dimensional state-space with $m$ input signals. For a solution to exist and be unique according to the Picard-Lindelöf theorem, we must choose activation functions such that
\begin{enumerate}
    \item $f_\theta$ is Lipschitz in $z$ and
    \item $f_\theta$ is continuous in $t$.
\end{enumerate}

The vanilla neural ODE is a first-order differential equation incapable of modeling higher order systems. Thus, for systems of order greater than one, we lift the state to include derivatives $x^{(1)}, x^{(2)}, \ldots $, forming an augmented state 
\begin{equation}
\begin{aligned}
    z &= \begin{bmatrix}
        x_1^{(0)} \\
        \vdots  \\
        x_n^{(o_n)}
    \end{bmatrix}.
\end{aligned}
\end{equation}
Here, $x_i^{(j)}$ denotes the $j$-th time derivative $\frac{d^jx}{dt^j}$ of the $i$-th state variable, up to order $o_i$. The augmented state $z$ with outputs $y$ evolves according to
\begin{equation}
    \begin{aligned}
    \dot{z} &= Az + Bf_\theta(z) \\
    y &= Cz\\
    z(0) &= z_0, 
    \end{aligned}
\end{equation}
where $A$ is the canonical companion matrix, $B$ is an input mask and $C$ is the output vector. The higher order initial values can either be estimated during training or specified a priori.

The neural ODE is then efficiently trained using the adjoint sensitivity method \cite{pontryagin2018mathematical} on some data-loss function at the sampled observations $y$ and and predictions $\hat{y}$
\begin{equation*}
    \mathcal{L}(\theta) = \ell(\hat{y}(\theta), y).
\end{equation*}

\begin{algorithm}[t]
    \caption{EstimateNeuralSurrogate}\label{alg:EstimateUDE}
    \begin{algorithmic}[1]
        \Statex \textbf{Input:} $\mathcal{D} = \{\tau\}_{i=1}^k$
        \Statex \textbf{Output:} $\hat{\mathcal{S}}$
        \State Initialize $\theta$
        \Repeat 
            \State $\hat{x}_i(t) \gets \text{ODESolve}(f_h, x_0^{(i)}, u^{(k)})$ 
            \State $L \gets  \sum_{i=1}^k\sum_{t=0}^T \ell \left ( y_t^{(i)}, x^{(i)}(t; \theta) \right )$
            
            \State Compute $\nabla_\theta L$ by the adjoint state method.
            \State $\theta \gets \theta - \eta \nabla_\theta L$
        \Until{convergence}
    \end{algorithmic}
\end{algorithm}
The adjoint sensitivity method works by first solving the dynamical system forward in time followed by computing the adjoint state $a(t_N) = \frac{d \mathcal{L}}{dx(t_N)}$ at the end of the solution interval. This adjoint state at $t_N$ serves as the initial condition for the adjoint differential equation solved backwards in time. Solving the adjoint differential equation can be done in a single ODESolve, only scaling with the number of function evaluations required by the integrator. 

The MLP parameters $\theta$ are then updated iteratively using gradient descent until some convergence criterion is met, as shown in Algorithm \ref{alg:EstimateUDE}. We begin training with ADAM and switch to a quasi-Newton method once ADAM satisfies a predefined convergence criterion. The quasi-Newton stage uses an approximate Hessian to accelerate convergence toward a local optimum, offering faster refinement than ADAM alone. To prevent the neural surrogate from bias accumulation towards trajectories collected early, we retrain it from scratch on the entire dataset $\mathcal{D}_k$ in Algorithm \ref{alg:learning-based falsification}, for each iteration $k$.

Neural ODEs can also be trained in parallel, scaling constantly with the number of data, as each ODESolve runs independently from the other trajectories. This also enables estimation with multiple shooting \cite{turan2021multiple}, by partitioning trajectories into sub-trajectories with initial conditions at some of the observations. 

A beneficial side-effect of estimating the dynamics as a neural ODE is that it provides continuous numerical derivatives. This is particularly useful in settings with measurement noise, where finite differencing or dynamic mode decomposition, used in Koopman theory, suffers inaccuracies. We emphasize the seamless ability to incorporate known dynamics and physical laws into the hybrid neural ODE and loss function, respectively, which reduces training time and improves generalization outside of training data. 

\subsection{Symbolic Distillation}
With the neural ODE trained in accordance to Algorithm ~\ref{alg:EstimateUDE}, we now estimate a symbolic surrogate using SR techniques \cite{la2021contemporary, makke2024interpretable}. This is done by sampling new derivative data from the trained surrogate and minimize the error between the symbolic expression and data. This is particularly useful when the outputs are noisy, as many SR methods are sensitive to noisy and sparsely sampled data. The surrogate dynamics of the neural ODE are continuous and smooth, which allows for smooth derivatives at arbitrary small sampling steps, aiding SR fitting. Sampling derivative data from more areas of the state space than just the training-data is of particular importance during the falsification stage, as the dynamics in the SQP solver should be an accurate approximation of the neural surrogate over the entire domain.

\begin{algorithm}[t]
    \caption{DistillSymbolic}\label{alg:EstimateSymb}
    \begin{algorithmic}[1]
        \Statex \textbf{Input:} $\mathcal{D} = \{\tau\}_{i=1}^k$ , $\hat{\mathcal{S}}$
        \Statex \textbf{Output:} $\hat{\mathcal{S}}_{symb}$
        \State Generate $\dot{\mathbf{z}}, \mathbf{z}$ from $\hat{\mathcal{S}}$ and $\mathcal{D}$
            \State $f^*(\mathbf{z}) \gets \arg \min_{f \in \mathcal{F}} \frac{1}{N}\sum_{i=1}^N(\dot{\mathbf{z}} - f(\mathbf{z}_i))^2$ 
            \State $\hat{\mathcal{S}}
            _{symb} \gets \arg \min_{f \in f^*} \mathcal{L}(\text{ODESolve}(\mathbf{z_0}, f), \mathbf{z})
            $
            \State \textbf{Return:} $\hat{\mathcal{S}}_{symb}$
            
    \end{algorithmic}
\end{algorithm}

Our proposed method is agnostic to the specific SR mechanism of choice, but some options include sparse identification of nonlinear dynamics (SINDy) \cite{brunton2016discovering} and genetic algorithms \cite{cranmerInterpretableMachineLearning2023}.

As a last step in SR distillation, we select the symbolic candidate which yields the lowest trajectory MSE-loss integrated over the horizon in $\mathcal{D}$. This avoids selecting a candidate which is over-fitted to derivatives instead of integration and regularizes against overly complex candidate expressions.

Using SR sparsifies the dynamics encoded by an otherwise dense MLP in the neural ODE into something symbolic. The main benefit of sparsifying the dynamics is to enable the use of sparse nonlinear optimizer (SNOPT) \cite{gill2002} and circumvent numerical gradient approximation in the following optimal control problem. 

\subsection{Optimal control synthesis}
Given a symbolic approximation of the surrogate dynamics, the STL specification can be encoded as a nonlinear program, and SNOPT can then be used to solve the resulting optimal control problem to search for a counterexample. Access to a symbolic expression of the dynamics enables exact analytical gradients of the surrogate, which are both faster and more accurate than numerically approximated gradients. In addition to being computationally cheaper, analytical derivatives avoid the noise and sensitivity that numerical differentiation introduces, particularly in non-smooth objective landscapes.

Minimizing the robustness of a specification is equivalent to maximizing the robustness of its negation, yielding the optimal control formulation for the discretized dynamics
\begin{equation}
\begin{aligned}
    \max_{\mathbf{x}, \mathbf{u}} & \: \rho^{\lnot \varphi}(y_0, y_1, \ldots, y_T) \\
    \text{s.t.} \: & x_0 \: \text{fixed} \\
    & x_{k+1} = f(x_k, u_k) \\
    & y_k = g(x_k, u_k) \\
    & u_{min} \leq u_k \leq u_{max}, \quad \forall k \in [0, \ldots, T]\\
    & x_{min} \leq x_k \leq x_{max}, \quad \forall k \in [0, \ldots, T].
\end{aligned}
\end{equation}

With a finite set of decision variables $u_0, \ldots, u_T$ and states $x_0, \ldots, x_T$, the optimal control problem is no longer a standard IVP that can make use of arbitrary ODE solvers. Instead, the dynamics must be transcribed using direct collocation methods. 
Furthermore, the decision variables are fixed to a set of predefined discrete time steps, which prevents the use of adaptive time stepping. Adding the step sizes $\Delta t_i = t_i - t_{i-1}$ as decision variables would also lead to spurious counterexamples, because the solver may exploit higher $\Delta t_i$ if it falsifies the specification, regardless of whether or not such behavior is consistent with the learned dynamics.

This lack of adaptive time stepping and the absence of intermediate evaluations of $f$ make the optimal control problem sensitive to stiffness in the dynamics. Moreover, the number of constraints required to encode the specification grows poorly with the depth of nested temporal operators.\footnote{For instance, a nested $\square_{[0, T]} \lozenge_{[0, T]} \varphi$ scales with $\mathcal{O}(T^2)$ compared to $\varphi$.} Thus, choosing an appropriate collocation method is crucial.


To make the STL robustness fully differentiable, we must also make smooth approximations of the $\min$ and $\max$ operators in the quantitative semantics. We do this in accordance to \cite{gilpin2020smooth, pant2017smooth}, by applying the log-sum-exponential (LSE) approximation:
\begin{equation}
    \begin{aligned}
        \max([a_1, ..., a_m]^T) & \approx \frac{1}{k} \log \left ( \sum_{i=1}^m e^{ka_i}\right ) \\
        \min([a_1, ..., a_m]^T) & \approx -\frac{1}{k} \log \left ( \sum_{i=1}^m e^{-ka_i}\right ).
    \end{aligned}
\end{equation}
Here, $k$ is a smoothing parameter for which the approximations become exact as $k \rightarrow \infty$, for a vector of discrete signal values $a_1, \ldots, a_m$. Such a robustness approximation does not change the non-convex nature of the problem, but local optima can now be solved for with sequential quadratic programming (SQP) solvers, such as SNOPT. The robustness function being asymptotically complete with respect to the specification 
entails that potential serendipitous solutions - falsifying the surrogate but not the SUT - only can arise from discretization errors, mislearnt surrogate or a too small smoothing coefficient.
\section{EXPERIMENTAL SETUP}
The implementation of our method was developed in the Julia programming language \cite{bezanson2017julia}. We use SymbolicRegression.jl \cite{cranmerInterpretableMachineLearning2023} for symbolic regression. The optimal control problem is formulated using Drake \cite{drake} and the STL specifications are smoothly approximated and encoded using STLpy \cite{kurtz2022mixed}. 
Lastly, we use Breach \cite{donze2010breach} with max semantics for robustness evaluation. All experiments and computations were carried out on a 13th Gen Intel(R) Core(TM) i7-13700H machine with 32GB memory. The framework is evaluated using the neural network controlled levitating magnet benchmark system \cite{MathWorksNARMA-L2}. The system is a feedback-controlled magnet levitating above a coil in which an electromagnetic field is induced. The inducing current is given from a NARMA-L2 feedback controller tasked with positioning the magnet at a reference value, bound to the interval $1 \leq \mathit{Ref} \leq 3$. The most important hyperparameters for our framework are summarized in Table \ref{tab:experiment-params}.

\subsection{Learning}
We initialize the data by parameterizing the input signal and sampling it using a corners-random strategy. The corners-random strategy works by randomly selecting one of the two bounds at each control point. This method has empirically shown to quickly falsify simpler specifications, potentially falsifying the specification already on initialization of the data \cite{ramezani2021testing}. We use equally long segments of 5 seconds and adhere to the upper and lower bounds $1 \leq \mathit{Ref} \leq 3$ for NN$_1$ and NN$_2$, and  $1.95 \leq \mathit{Ref} \leq 2.05$ for NN$_x$, respectively.
The SUT experiment signals are sampled at a constant sampling time of $0.01$ seconds, according to the instructions in the ARCH-COMP 2024 competition.


To ensure a fair comparison with other tools, no a priori assumptions of the dynamics are made, setting $f_k = 0$. Solving the IVPs is done with the Tsit5 solver, which is a 4th-order Runge-Kutta solver.

\begin{table}[h]
\centering
\footnotesize
\setlength{\tabcolsep}{2pt}
\caption{Overview of the hyperparameters used in our falsification experiments.}
\begin{tabular}{l l|l l|l l}
\toprule
\multicolumn{2}{c|}{\textbf{Learning}} &
\multicolumn{2}{c|}{\textbf{Symbolic Regression}} &
\multicolumn{2}{c}{\textbf{Falsification (OCP)}} \\ \hline
Learning rate        & $5 {e}{-2}$   & Complexity & $< 30$              & \makecell[l]{Smoothing \\ parameter} & 2 \\ \hline
\makecell[l]{Epochs \\ (ADAM)}             & 300                  & Iterations           & 100                 & \makecell[l]{Experiment \\ budget}  & 10 \\ \hline
\makecell[l]{Iterations \\ L-BFGS}       & 20                   & Population size           & 50                  & Optimizer                & SNOPT \\ \hline
\makecell[l]{Hidden neurons}             & 16, 8        & Unary operators           & exp, sin, cos       & Collocation $\Delta t$            & $0.2$s, $0.1$s  \\ \hline
Loss function             & MSE                  & Binary operators          & $+,-,\times,\div$   &  &  \\ \hline
System order              & 2                    &                           &                     &    &  \\
\bottomrule
\end{tabular}
\label{tab:experiment-params}
\end{table}

For the neural network controlled levitating magnet, we use a fully connected MLP for second order dynamics with one neuron per initial state and input signal $f: \mathbb{R}^n \times \mathbb{R} \mapsto \mathbb{R}^n$, with two hidden layers consisting of $16$ and $8$ hidden neurons, respectively. For activations we use $\tanh$ functions in all layers except the output layer, which satisfies the Picard-Lindelöf conditions for uniqueness and existence of a solution to the initial-value problem (\ref{eq:IVP}), due to $\tanh$ being Lipschitz.

No physics loss is incorporated for the benchmarks, implying a pure data-loss from the observations. We use the mean-squared error (MSE) to guide learning:
\begin{equation}
    \ell(\theta) = \frac{1}{T}\sum_{t=0}^T (\hat{y}_t(\theta) - y_t)^2.
\end{equation}
For training, a fixed number of epochs is set, instead of relying on early-stopping techniques. This is done to avoid spending experiments on acquiring costly validation data that does not explicitly improve training. We selected the learning rate of $5\mathrm{e}{-2}$ based on preliminary evaluations, which showed training instability for higher learning rates and slow convergence for lower values. Similarly, increasing the number of epochs beyond $300$ did not display significant improvements, in particular for early iterations, when $\mathcal{D}$ contains only a few trajectories.

\subsection{Symbolic Distillation}
We use SymbolicRegression.jl \cite{cranmerInterpretableMachineLearning2023} for SR, which is an open-source Julia package for discovering analytical expressions of functions, from data. The method encodes the computation tree of a solution as a genetic individual and combines evolutionary with gradient-based optimization to arrive at an optimally fit computation tree. The search space of possible symbolic expressions is limited to have complexities less than 30. Preliminary results showed that both trajectory and derivative losses started to plateau around that value. 
To keep the function space $\mathcal{F}$ general, we search for functions broadly, incorporating the unary operators $\{\exp(\cdot), \sin(\cdot), \cos(\cdot)\}$ and binary operators $\{+, -, (\cdot), \div \}$. We strategically omit operations which can yield undefined computations over the domain, such as $\log(\cdot)$ and $\sqrt{\cdot}$. Since our framework cannot guarantee that SNOPT stays in the definition set of $f$, it risks crashing if such operators are included.

The symbolic distillation step is solved for 100 iterations using 50 populations. The symbolic candidate is selected greedily by minimum trajectory loss. Candidate dynamics that are not driven by inputs, i.e., expressions for which $\frac{\partial f}{\partial u} = 0$ or expressions that are numerically unstable during simulation, are also filtered out.

\subsection{Optimal Control Falsification}
We set the experiment budget of SUT to 10. If a counterexample on the SUT is not found by the 10th iteration, the experiment is considered unsuccessful. Robustness evaluations on the SUT are done using Breach \cite{donze2010breach}. 

As shown in Table \ref{tab:specs}, the specifications NN and NN$_{\beta=0.04}$ both have a triple nesting of temporal operators, with the outermost spanning almost the entire time horizon. To reduce the number of equality constraints, we use coarser discretization scheme for said specifications, setting $\Delta t = 0.2$s. NN$_x$ has a much shorter horizon and less nested temporal operators which allows a a finer discretization $\Delta t = 0.1$s. All discretizations are made with equidistant steps. A coarser step size introduces larger truncation and global errors, in particular if using a explicit Euler discretization scheme.

\begin{table}[t]
    \centering
    \caption{The specifications used in the levitating magnet benchmark problem.}
    \begin{tabular}{|c|c|} \hline
        Spec. & STL definition \\ \hline
        NN &  \makecell{$
            \begin{aligned}
            \Box_{[1, 37]} \Big ( (|\mathit{Pos} - \mathit{Ref} | > 0.005 + 0.03|\mathit{Ref}|) \\ \longrightarrow \lozenge_{[0,2]} \Box_{[0, 1]}  \lnot \left ( 0.005 + 0.03|\mathit{Ref}| \leq |\mathit{Pos} - \mathit{Ref} \:| \right ) \Big )
            \end{aligned}
        $} \\ \hline
        NN$_{\beta = 0.04}$ & \makecell{$
            \begin{aligned}
            \Box_{[1, 37]} \Big ( (|\mathit{Pos} - \mathit{Ref} | > 0.005 + 0.04|\mathit{Ref}|) \\ \longrightarrow \lozenge_{[0,2]} \Box_{[0, 1]}  \lnot \left ( 0.005 + 0.04|\mathit{Ref}| \leq |\mathit{Pos} - \mathit{Ref} \:| \right ) \Big ) \\ 
            \end{aligned}$} \\ \hline
            NNx   & \makecell{$
            \begin{aligned}
            & \lozenge_{[0, 1]} (Pos > 3.2) \\ 
            & \land \lozenge_{[1, 1.5]} (\square_{[0, 0.5]} (1.75 < Pos < 2.25)) \\ 
            & \land \square_{[2,3]} (1.825 < Pos < 2.175) \\
            & 1.95 \leq \mathit{Ref} \leq 2.05
            \end{aligned}
        $} \\ \hline
    \end{tabular}
    \label{tab:specs}
\end{table}

To counter this, we assume piece-wise linear dynamics $f$ in each interval $t \in [t_k, t_{k+1}]$ with interval length $h_k = t_{k+1} - t_k$. This is the direct collocation using the trapezoidal method \cite{kelly2017introduction}:
\begin{equation}
    f(t) = \dot{\mathbf{z}}(t) \approx f_k + \frac{t - t_k}{h_k}(f_{k+1} - f_k),
\end{equation}
where $f_k = f(z_k, u_k)$.
This produces a quadratic spline interpolation for the state trajectory in the segment:
\begin{equation}
    x(t) \approx x_k + f_k \cdot (t - t_k) + \frac{(t-t_k)^2}{2h_k}(f_{k+1} - f_k),
\end{equation}
which at $t = t_{k+1}$ yields
\begin{equation}
    x_{k+1} \approx x_k + \frac{h_k}{2}(f_{k+1} + f_k).
\end{equation}
This equality constraint is repeatedly applied between each pair of collocation points, encoding the discretized trajectory as a chain of equality constraints.

The local and global truncation errors of the trapezoidal collocation scale with $\mathcal{O}(h_k^3)$ and $\mathcal{O}(h_k^2)$, respectively. This can be leveraged to make larger steps, effectively reducing the number of decision variables and equality constraints for large specifications, or maintain higher accuracy for specifications that are difficult to falsify.

Furthermore, we warm-start the optimal control problem by setting the initial guess of states and inputs as a solution to the IVP of the symbolic surrogate. This assures that equality constraints are satisfied before running the solver, instead of wasting iterations in the solver to satisfy them. It also reduces the risk of initializing the decision variables in poor local optima. To ensure termination of SNOPT, we set a major iterations limit of $2000$.

The optimal control problem is solved using a modified version of STLpy \cite{kurtz2022mixed}, used for approximating smooth robustness and pyDrake \cite{drake} to interface the SNOPT solver. We set the smoothing parameter to moderate value $k=2$ for the LSE approximation, to balance gradient smoothness and stiffness.

For fair methodological comparison to the tools available in \cite{khandait2024arch}, we separate fluke counterexamples from converged counterexamples in the optimal control problem. A fluke counterexample falsifies the SUT, despite building on an optimal control solution that failed to converge. As this is an artifact rather than a trusted solution, we add the SUT trajectory to $\mathcal{D}$, but do not consider it as a valid counterexample.
\section{RESULTS}
We apply our method on a subset of the ARCH-COMP 2024 benchmark problems \cite{khandait2024arch} for evaluation, namely all specifications related to the neural-network controlled levitating magnet \cite{MathWorksNARMA-L2}. The results are compared to other state-of-the-art methods from the competition, namely ARIsTEO, ATheNA, EXAM-Net, ForeSee, FReaK, Moonlight, OD, $\Psi$-TaLiRo. The performance metrics for the other tools are retrieved from the ARCH-COMP 2024 paper and are not reevaluated or confirmed by the authors of this paper. We specifically omit the tools NNFal and FalCAuN, as these do not have any associated performance metrics reported on the levitating magnet specifications.

The ARCH-COMP 2024 benchmark contains two different instances. The instances parameterize the input signal differently but share common specifications. Instance 1 allows arbitrary input signals to be designed by participants of the competition, whereas instance 2 constraints the input to be parameterized according to some predetermined specification as described in \cite{khandait2024arch}. However, to demonstrate our framework's agnosticism to input-signal parameterization, we let the optimal control problem construct the piecewise constant input-signal freely within bounds, and compare it against both instances.

The benchmark problem we use for evaluation consists of a levitating magnet, controlled by a neural network feedback controller. It is the only benchmark system from the ARCH-COMP 2024 problem set with multiple STL specifications and no mode-switching or mixed discrete-continuous states. It has three different specifications shown in Table \ref{tab:specs}, all related to setpoint tracking. Specifications NN and NN$_{\beta=0.04}$ describe a convergence behavior of $\mathit{Pos}$ towards $\mathit{Ref}$ with some tolerance factor $\beta$. NN$_x$ describes, in more detail, the settling process during the first three seconds as a conjunction of three conjuncts. 

Due to the algorithm’s stochastic nature, we run the method $10$ times for each benchmark problem and specification and report aggregate performance metrics. We evaluate performance based on three main metrics: average and median number of experiments needed to falsify a given specification, $\bar{S}, \: \tilde{S}$, as well as the falsification rate (FR). The FR describes how many of the reiterations manage to find a counterexample within a maximum iterations count. 

\subsection{Learning}
Training the neural ODE surrogate accurately requires trajectory data that sufficiently cover the relevant regions of the state space. If trajectories are missing from areas where the specification is likely to be violated, both the neural and symbolic approximations in those regions may be inaccurate. Consequently, the optimal control problem may produce a spurious falsifying trajectory that does not correspond to a true violation of the SUT.

Avoiding overfitting presents an additional challenge when training the neural ODE. Standard approaches rely on cross-validated early stopping, triggered when the validation loss begins to increase while the training loss continues to decrease. In our setting, however, the number of available trajectories is directly linked to the algorithm’s performance, rendering such early stopping strategies potentially counterproductive—especially when $\mathcal{D}$ contains only a few trajectories.

A benefit of neural ODEs is that they approximate a smooth vector field which enables local extrapolation of the learned dynamics into neighboring areas of the covered state space. When spurious counterexamples are found to the falsification problem, they are added to the training data, expanding the part of state space in which the surrogate dynamics are accurate after further learning.

\subsection{Symbolic Distillation}
\begin{figure}[t]
    \centering
    \includegraphics[width=1\linewidth]{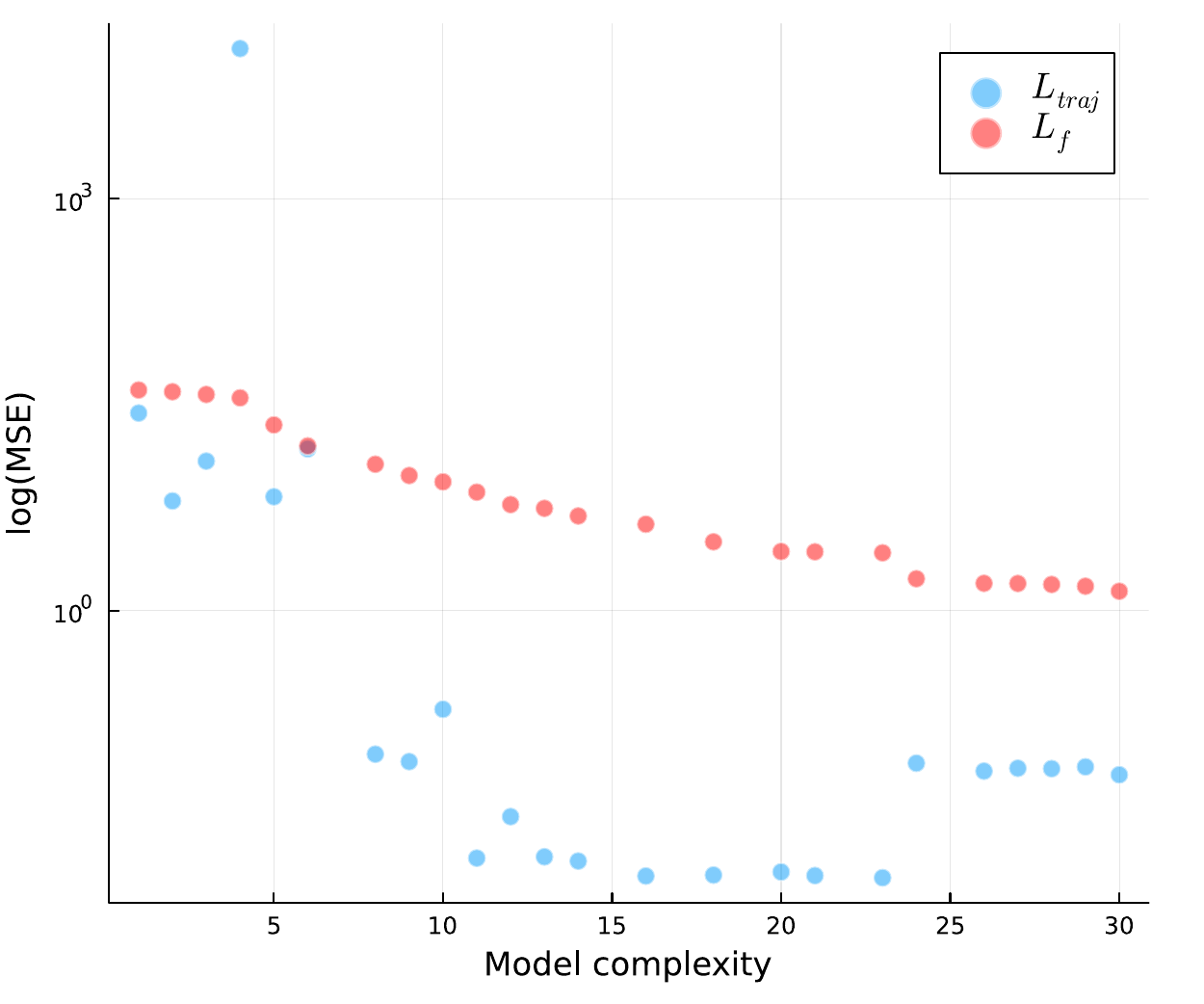}
    \caption{Blue points indicate the MSE trajectory loss $\mathbf{L_{traj}}$, computed by simulating each candidate model along trajectories in $\mathcal{D}$. Red points represent the MSE loss in dynamics. Notably, a candidate model that minimizes the dynamics error in the sampled points does not necessarily achieve minimal error when used in trajectory simulation.}
    \label{fig:complexities}
\end{figure}

The complexity of a symbolic regression candidate is a heuristic measurement of how many terms and operators are included. As shown in Figure \ref{fig:complexities}, the SR loss is generally reduced by increasing the complexity of an expression. However, there exists a discrepancy between the minimal derivative fitness of a candidate and the trajectory loss for said candidate. In the figure, the ideal candidate from SR would be of complexity 30, but that same candidate is outperformed by a less complex candidate, namely 23, when simulating a full trajectory instead and comparing it with data from $\mathcal{D}$.
For symbolic candidate selection in the neural network benchmark, we find that trajectory fitness improves with higher complexities before flattening out around complexity 25. Although higher complexities become costlier to evaluate in the SNOPT solver, we limit the candidate selection strategy to the one described. 

\begin{figure}[t]
    \centering
    \includegraphics[width=1.0\linewidth]{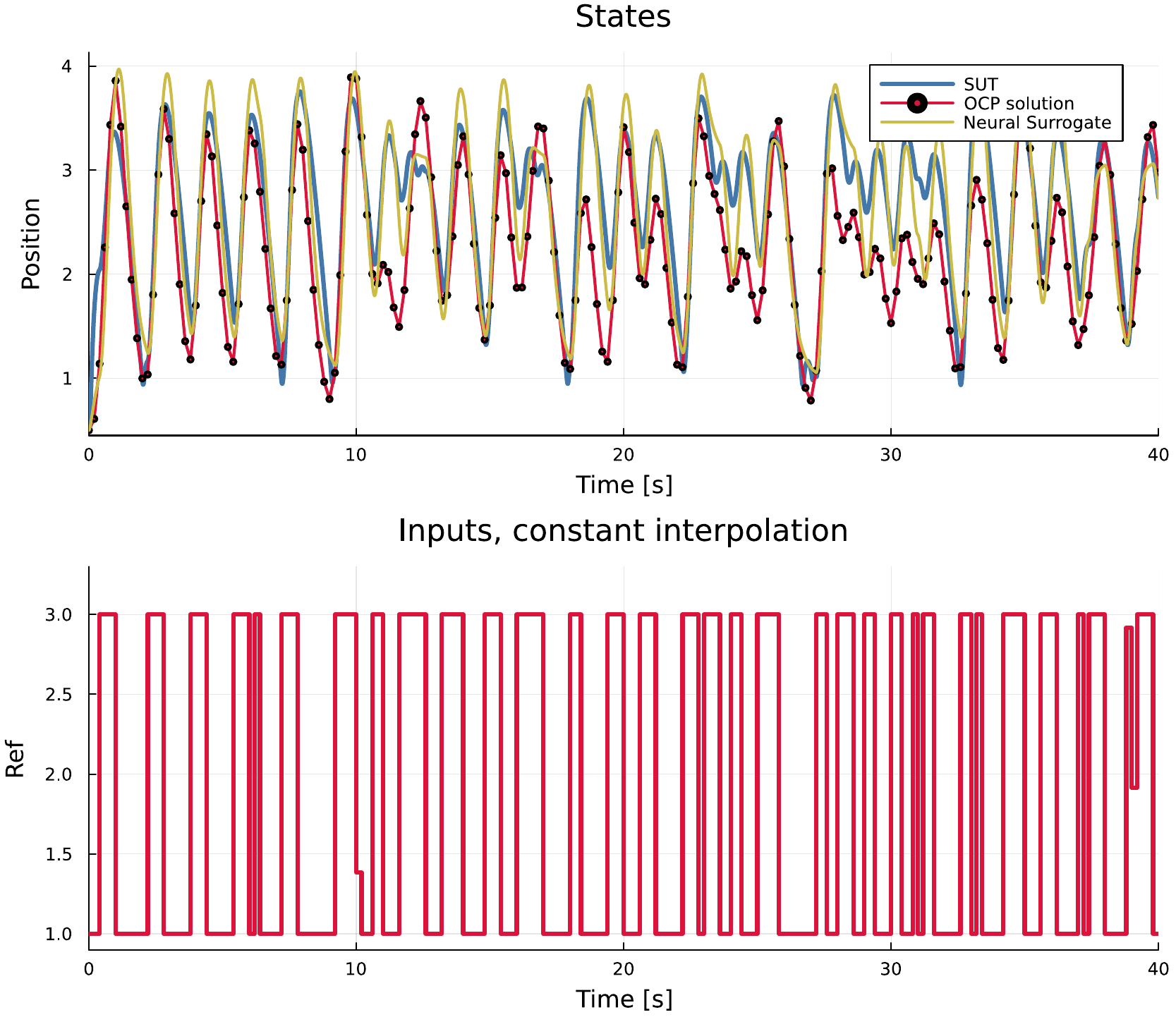}
    \caption{Synthesized counterexample together with the corresponding SUT trajectory, symbolic and neural surrogates. The resulting SUT robustness is -2.014 and the robustness from the optimal control problem is -2.13.}
    \label{fig:falsification_example}
\end{figure}

Although the distillation step in general introduces additional approximation errors to the surrogate, we find that the distilled symbolic surrogate accurately represents its neural counterpart in the regions relevant to falsification, as shown in Figure \ref{fig:falsification_example}. Local deviations occur but tend to have minor qualitative impact on the trajectory over the entire time interval.

\subsection{Falsification}

SNOPT is a local optimizer that seeks to minimize the objective function. In the context of falsification, however, a solution may still be useful even if it is suboptimal with respect to robustness minimization. As a result, solutions that do not fully minimize the objective can still end up falsifying the SUT. In cases where the optimal control problem fails to find a locally optimal solution, the resulting suboptimal trajectory may still yield a negative robustness on both the SUT and the surrogate. We discard such incidental successes for counterexample purposes, as they arise from serendipitous falsifications, rather than from the intended operation of our method. Nonetheless, we add these serendipitous trajectories resulting from performing experiment of the SUT to the training data.




Table \ref{tab:combined_transposed} shows the resulting aggregate performance metrics over repeated executions, where each execution corresponds to one benchmark problem and specification. We repeat the falsification 10 times on each specification, to reduce the influence of stochasticity and randomness in the algorithm. The results show that our approach requires roughly as many or fewer SUT experiments compared to other tools to find counterexamples on the selected benchmarks. Across repeated runs, both NN and NN$_{\beta=0.04}$ are consistently falsified within at most four iterations of the algorithm. When a counterexample is not found from solving the optimal control problem during the first iteration of Algorithm \ref{alg:learning-based falsification}, it typically means that SNOPT failed to produce a solution within the default termination criteria. These failures are likely due to the highly nonlinear and coupled nature of the dynamics and STL constraints.

\begin{figure*}[t]
    \centering
    \includegraphics[width=1\linewidth]{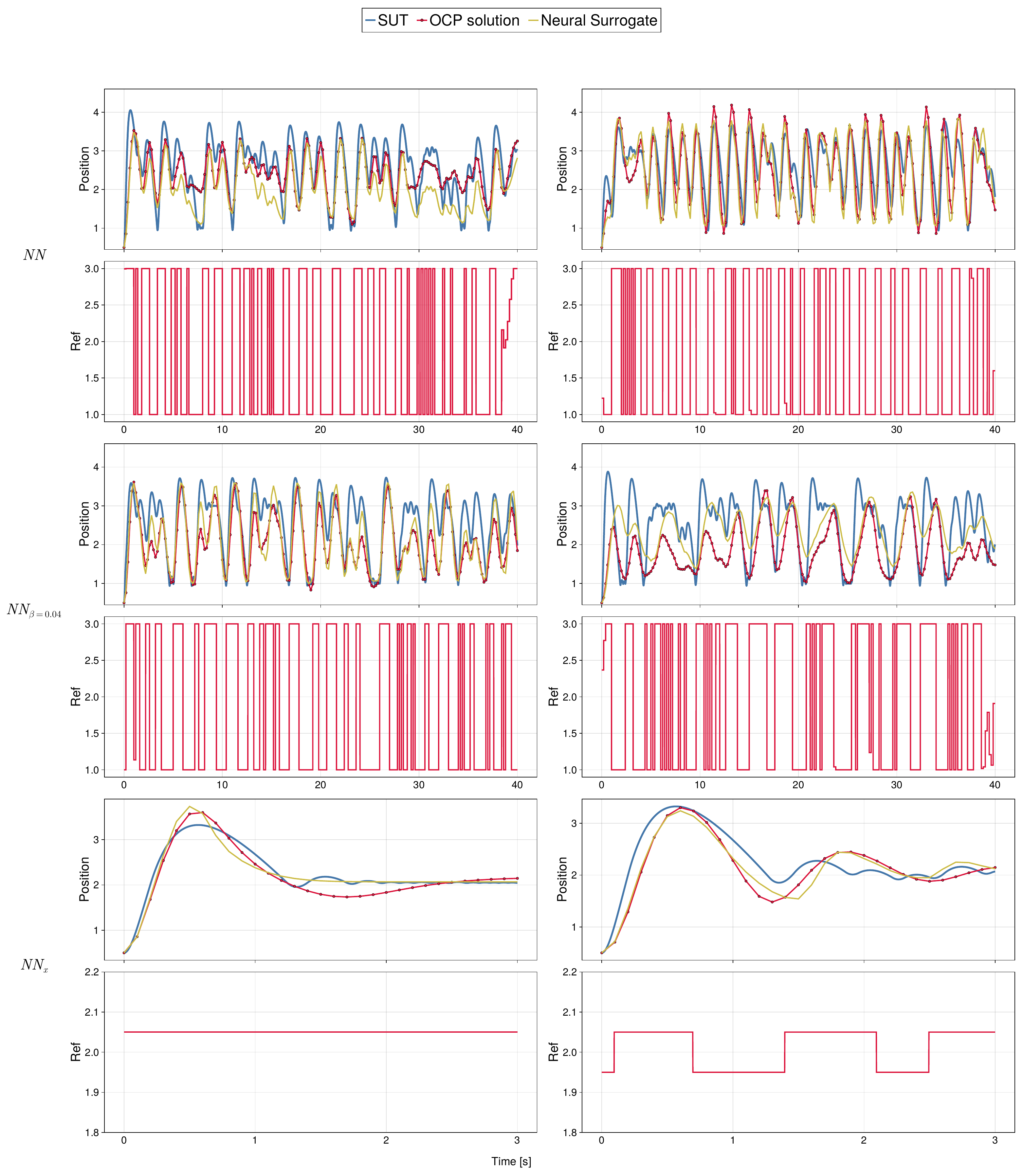}
    \caption{Results from two independent falsification executions per specification, using our framework. The SUT (blue) is executed using the candidate counterexample from the optimal control problem (red). The neural surrogate (yellow) is plotted for comparison. All results show a falsifying trace except the bottom leftmost. For specifications NN and NN$_{\beta = 0.04}$, the violation is achieved by frequently altering the reference value, preventing a convergence of the position to the reference.} 
    \label{fig:instance1}
\end{figure*}

The resulting counterexamples pursue an input signal whose reference value changes rapidly, as illustrated in Figure \ref{fig:instance1}. This behaviour is expected, since repeatedly shifting the reference causes the system to violate the requirement that the actual position must settle within two seconds after diverging from the reference, thereby minimizing robustness.

First, we observe that the neural surrogate captures the qualitative dynamics well, with it's trajectories staying close to the validated SUT counterexample. Second, the symbolic distillation is largely successful: it reproduces the local dynamics accurately and introduces only minor integration errors. The optimal control solution also satisfies the equality constraints to a high degree. The most common cause behind requiring more than one iteration is that the optimal control problem terminates unsuccessfully.


\begin{table}[t]
\centering
\caption{Combined performance comparison for ARCH-COMP 2024 instances 1 and 2 with our framework. The metrics are averaged over 10 repeated experiments for each specification and instance. Metrics: number of successful falsifications FR out of 10, mean $\bar{S}$, and median $\tilde{S}$ number of experiments required to find a counterexample. Dashes imply that no counterexamples were found within the specified experiment budget in the competition. Missing entries mean that the data is unavailable. To evaluate the framework, a budget of 10 experiments was used.}
\resizebox{.5\textwidth}{!}{
\begin{tabular}{l|ccc|ccc}
\hline
 & \multicolumn{3}{c|}{\textbf{Instance 1}} & \multicolumn{3}{c}{\textbf{Instance 2}} \\
\textbf{Framework} &
FR & $\bar{S}$ & $\tilde{S}$ &
FR & $\bar{S}$ & $\tilde{S}$ \\
\hline
\multicolumn{7}{c}{\textbf{NN}} \\
\hline
Our Approach   & 10 & 3.0 & 4.0 & 10 & \textbf{3.0} & \textbf{4.0} \\
UR             & 10 & 38.6 & 27.5 & 10 & 277.2 & 158.5 \\
ARIsTEO        & 5 & 133.4 & 108.0 & 10 & 119.2 & 91.0  \\
ATheNA         & 9 & 386.0 & 276.0 & 10 & 49.7 & 47.5 \\
EXAM-Net       & 10 & 47.2 & 47.0 & 10 & 47.2 & 47.0  \\
FORESEE        & 10 & 122.8 & 97.5 & 10 & 122.8 & 97.5 \\
FReaK          & 10 & \textbf{2.0} & \textbf{2.0} & 10 & 37.2 & 25.0  \\
Moonlight      & 10 & 26.8 & 22.0 & 10 & 48.0 & 32.0  \\
OD             & 10 & 53.0 & 39.0 & 10 & 53.0 & 39.0  \\
$\psi$--TaLiRo & 10 & 366.9 & 236.0 & 10 & 75.9 & 83.0 \\
\hline
\multicolumn{7}{c}{\textbf{NN$_{\beta = 0.04}$}} \\
\hline
Our Approach   & 10 & \textbf{2.4} & \textbf{2.0} & 10 & \textbf{2.4} & \textbf{2.0}   \\
UR             &   &   &   &   &   & \\
ARIsTEO        & 0 & - & - & 1 & 55.0 & 55.0 \\
ATheNA         & 0 & - & - & 2 & 503.0 & 503.0 \\
EXAM-Net       &   &   &   &   &   & \\
FORESEE        & 6 & 963.7 & 1019.0 & 6 & 963.7 & 1019.0 \\
FReaK          & 10 & 30.9 & 33.0 & 10 & 534.0 & 472.0 \\
Moonlight      &   &   &   &   &   &   \\
OD             & 4 & 591.0 & 690.0 & 4 & 591.0 & 690.0 \\
$\psi$--TaLiRo & 0 & - & - & 0 & - & - \\
\hline
\multicolumn{7}{c}{\textbf{NN$_x$}} \\
\hline
Our Approach   & 6 & 4.8 & 3.0 & 6 & 4.8 & 3.0 \\
UR             &   &   &   & 8  & 457.1  & 380.5 \\
ARIsTEO        & 0 & - & - & 7 & 11.4 & 6.0 \\
ATheNA         & 0 & - & - & 9 & 182.6 & 178.0 \\
EXAM-Net       & 0 & - & - & 0 & - & - \\
FORESEE        &  &  &  &  &  &  \\
FReaK          & 10 & \textbf{192.3} & \textbf{165.5} & 10 & 165.3 & \textbf{65.5} \\
Moonlight      & 9 & 142.7 & 128.0 & 0 & - & - \\
OD             & 0  & -  & -  &  0 & -  & - \\
$\psi$--TaLiRo & 0 & - & - & 10 & \textbf{143.7} & 135.0 \\
\hline
\end{tabular}}
\label{tab:combined_transposed}
\end{table}

\section{CONCLUSIONS}
We have presented a novel falsification framework that integrates neural surrogate modeling, symbolic regression and optimal control to synthesize falsifying counterexamples. The method operates by iteratively training a neural ODE to approximate the SUT dynamics and distilling a symbolic representation suitable for optimal control synthesis. Having access this symbolic approximation of the SUT enables the use of a sparse nonlinear optimizer (SNOPT) to solve an optimal control problem, computing the least robust trajectories. Across three benchmark specifications from the ARCH COMP 2024 competition, the method is consistently able to find counterexamples to two specifications with roughly equal or fewer experiments than the tools from the competition. For the third specification - which only one other tool successfully falsifies across both instances - our framework only requires on average 4.8 experiments with a falsification rate of 6/10. 

While the overall iterative surrogate refinement loop is conceptually similar to recent recent work \cite{bak2024falsification, menghi2020approximation}, our approach differs fundamentally in how the system dynamics are modeled and how the falsifying counterexamples are computed. Modeling the state variables explicitly as a neural ODE enables learning of nonlinear dynamics as well as incorporating a priori knowledge of the dynamics, increasing learning speed and generalization beyond training data. Our framework does not necessitate a parameterization of the input signal, which typically requires expert insight. Instead, it treats the discretized input signal as a finite set of decision variables to be optimized. 

Future work includes extending the framework to support hybrid systems, with both switching conditions and mixed discrete-continuous states. 

\clearpage

\bibliographystyle{ACM-Reference-Format}
\bibliography{refs}

\end{document}